\documentclass[runningheads]{llncs}

\usepackage[colorlinks]{hyperref}
\usepackage{url}
\usepackage{amsmath}
\usepackage{algorithm}
\usepackage{algorithmic}

\usepackage{subfigure}
\usepackage{lipsum}
\usepackage{multirow}
\usepackage{booktabs}
\usepackage{graphicx}
\usepackage[normalem]{ulem}
\usepackage{amsmath}
\usepackage{amssymb}
\usepackage{mathrsfs}
\usepackage{float}
\usepackage[colorinlistoftodos]{todonotes}

\usepackage[misc]{ifsym}

\newcommand{\name}{HAFM}
\begin{document}

\title{HAFM: Hierarchical Autoregressive Foundation Model for Music Accompaniment Generation}
\titlerunning{\name}
%
\author{Jian Zhu\inst{1} \and
Jianwei Cui\inst{2} \and
Yunlong Xue\inst{1} \and 
Shihao Chen\inst{2} \and \\
Yubang Zhang\inst{2} \and
Linyao Yang\inst{1} \and
Cheng Luo\inst{3,*} \and
Jun Sun\inst{1,*}}

%
\institute{Zhejiang Lab \\
\email{\{qijian.zhu\}@outlook.com} \and 
University of Science and Technology of China \\
\email{\{jwcui\}@mail.ustc.edu.cn} \and
Zhejiang International Studies University \\ 
\email{\{luo\_cheng\}@zisu.edu.cn} 
\\}

\toctitle{HAFM}
\tocauthor{}
\authorrunning{Jian Zhu et al.}

%

%
%

%

\maketitle
\def\thefootnote{*}\footnotetext{Corresponding author.}

\begin{abstract}
Music accompaniment generation aims to automatically produce instrumental accompaniments that are rhythmically, harmonically, and timbrally coherent with a given vocal input, with broad applications in personalized music creation, arrangement assistance, and music education. Existing approaches, primarily operating in the symbolic domain or relying on single-stage audio generation frameworks, commonly suffer from insufficient high-level semantic structure modeling, limited acoustic detail reconstruction, and weak conditional controllability. To address these limitations, this paper proposes HAFM, a Hierarchical Autoregressive Foundation Model for vocal-conditioned music accompaniment generation. The model employs a dual-rate tokenization strategy in which $50$ Hz HuBERT semantic tokens capture high-level musical structure and $75$ Hz EnCodec acoustic tokens encode fine-grained acoustic content, enabling explicit disentanglement of semantic and acoustic representations. Building on this foundation, a three-stage cascaded generation framework is designed to progressively generate semantic tokens, coarse acoustic tokens, and fine acoustic tokens, refining the accompaniment from global structure to local detail. A classifier-free guidance mechanism is further incorporated to strengthen vocal conditioning during inference, while modern Transformer components including QK normalization, GEGLU activation, RMSNorm, and T5-style relative position bias are integrated to improve optimization stability and generation quality. Objective evaluation on the MUSDB18 dataset demonstrates that the full three-stage model achieves a Fr{\'e}chet Audio Distance (FAD) score of $1.71$, representing an $18.6\%$ relative improvement over the two-stage baseline (FAD = $2.10$). Subjective listening tests show that the generated accompaniments achieve a $51.5\%$ preference rate against ground-truth accompaniments in head-to-head comparisons, and substantially outperform the random baseline in terms of rhythmic alignment, harmonic compatibility, and overall musical coherence. The source code and demo are available at https://github.com/HackerHyper/HAFM.git.
\keywords{Vocal Accompaniment \and Music Generation \and Audio Language Model \and  Autoregressive Model}
\end{abstract}
\section{Introduction}

Singing is one of the most intuitive ways for people to engage in musical expression. It is not only a performance medium for interpreting existing works, but can also serve as a natural interactive interface for driving music creation—treating the human voice as an intuitive control signal that enables anyone who can sing to automatically generate personalized accompaniment. This has given rise to research on the vocal accompaniment generation task: given a separated vocal input, generate an instrumental waveform that can be mixed with the voice to produce coherent music. This task has broad application value in music education, entertainment, and assisted composition.

Music accompaniment generation often relies on neural audio codecs and audio language models. Neural audio codecs compress continuous waveforms into discrete token sequences and form the foundation of audio language models. Representative neural audio codecs include SoundStream~\cite{zeghidour2021soundstream}, which achieves high-quality audio compression at 16 kHz using Residual Vector Quantization (RVQ) with multi-scale STFT discriminators and reconstruction losses; EnCodec ~\cite{defossez2022encodec} extends this to 24 kHz and 48 kHz, introducing multi-scale discriminators and improved loss functions to enhance audio fidelity while maintaining low bitrates. Beyond acoustic codecs, semantic codecs such as HuBERT ~\cite{hsu2021hubert} learn self-supervised representations through masked prediction, capturing high-level semantic content of speech. Audio language models apply language modeling techniques to audio generation. Notable examples include AudioLM ~\cite{borsos2022audiolm}, which adopts hierarchical tokenization (semantic + acoustic) and cascaded Transformers to achieve high-quality audio continuation; MusicLM~\cite{agostinelli2023musiclm}, which extends this approach to text-conditioned music generation; and VALL-E~\cite{wang2023neural}, which demonstrates the effectiveness of neural codec language models for zero-shot speech synthesis.

Music accompaniment generation can be categorized into symbolic-domain and audio-domain approaches. Early accompaniment generation systems were primarily symbolic-domain methods—for example, MySong~\cite{simon2008mysong} generates chords and MIDI accompaniment from vocal melody, realizing voice-driven music accompaniment generation, but requires preprocessing steps such as pitch detection and symbolic conversion, and the generated MIDI timbres are limited. In recent years, advances in deep learning have driven research into audio-domain accompaniment generation. Progress has been made in instrument-specific audio-domain accompaniment generation: JukeDrummer~\cite{dhariwal2020jukebox} uses a Transformer VQ-VAE to generate drum track accompaniment, and BassNet~\cite{grachten2020bassnet} synthesizes rhythmically coherent bass tracks via a variational gated autoencoder. However, these methods target only a single instrument and cannot generate full multi-instrument accompaniment. SingSong~\cite{donahue2023singsong}is the first end-to-end audio-domain vocal accompaniment generation system, adopting the AudioLM framework with w2v-BERT~\cite{chung2021w2vbert} semantic features and the SoundStream codec. Nevertheless, SingSong relies on an encoder-decoder (T5~\cite{raffel2020exploring}) architecture, which may not fully exploit the autoregressive properties suited to generative tasks, and its use of 16 kHz SoundStream limits codec expressiveness, making it difficult to capture rich timbral detail.

To address these limitations, this paper proposes the Hierarchical Autoregressive Foundation Model (HAFM), advancing vocal accompaniment generation through three contributions: (1) Dual-rate codec tokenization. HuBERT semantic tokens at 50 Hz are used for vocal conditioning, while EnCodec acoustic tokens at 75 Hz are used for instrumental generation, yielding richer representations than single-codec approaches. (2) Three-stage hierarchical autoregressive architecture. Generation is decomposed into three stages—semantic, coarse acoustic (4 codebooks), and fine acoustic (4 codebooks)—each modeled by a decoder-only Transformer~\cite{vaswani2017attention} with Classifier-Free Guidance (CFG)~\cite{ho2021classifier}. (3) Modern Transformer design. Incorporating QK normalization ~\cite{dehghani2023scaling}, GEGLU activations ~\cite{shazeer2020glu}, RMSNorm~\cite{zhang2019root}, and T5 relative position bias~\cite{raffel2020exploring} to improve training stability and generalization to long sequences.

\begin{figure}
  \centering
  \includegraphics[width=9.5cm]{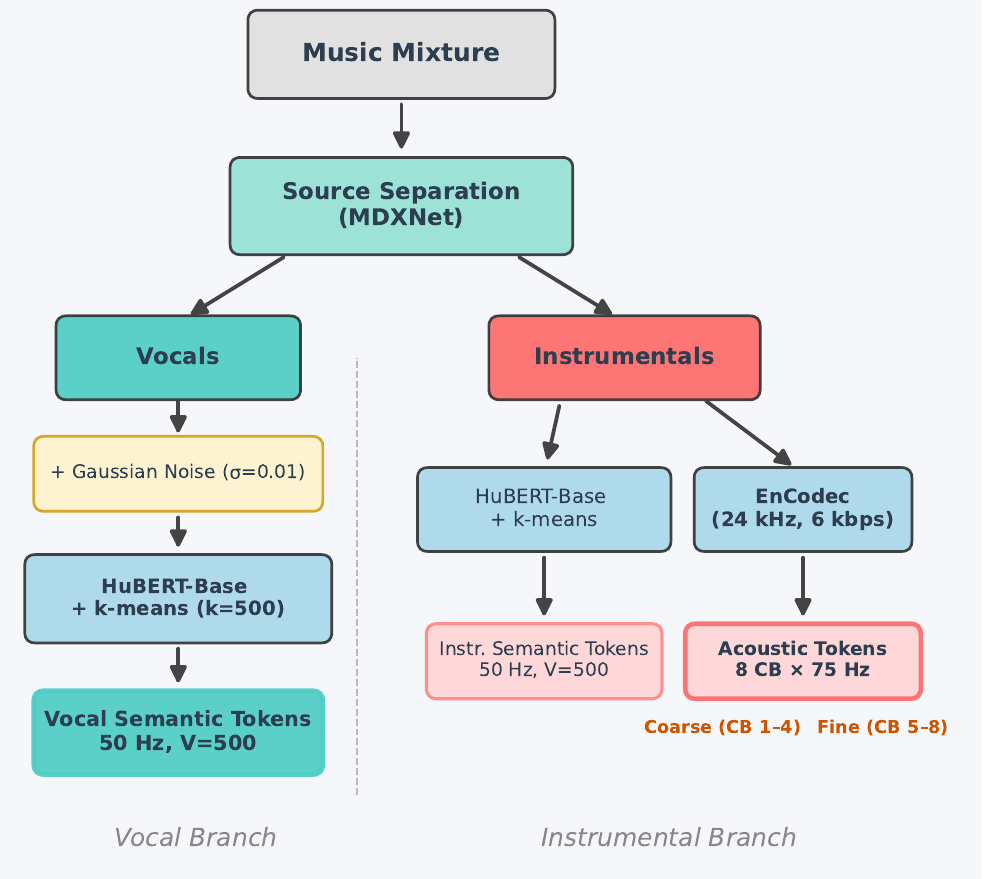}
  \caption{Data preprocessing pipeline. Source separation (MDXNet) extracts aligned vocal and instrumental pairs from music mixtures. Vocals are augmented with Gaussian noise ($\sigma\!=\!0.01$) and encoded via HuBERT-Large (layer~9) with $k$-means ($k\!=\!500$) into semantic tokens at 50\,Hz. Instrumentals are encoded into both semantic tokens (HuBERT, 50\,Hz) and acoustic tokens via EnCodec (8 codebooks at 75\,Hz), split into coarse (CB~1--4) and fine (CB~5--8) groups.}
  \label{fig:preprocess}
\end{figure}

\begin{figure*}[!t]
	\centering
	\includegraphics[width=12cm]{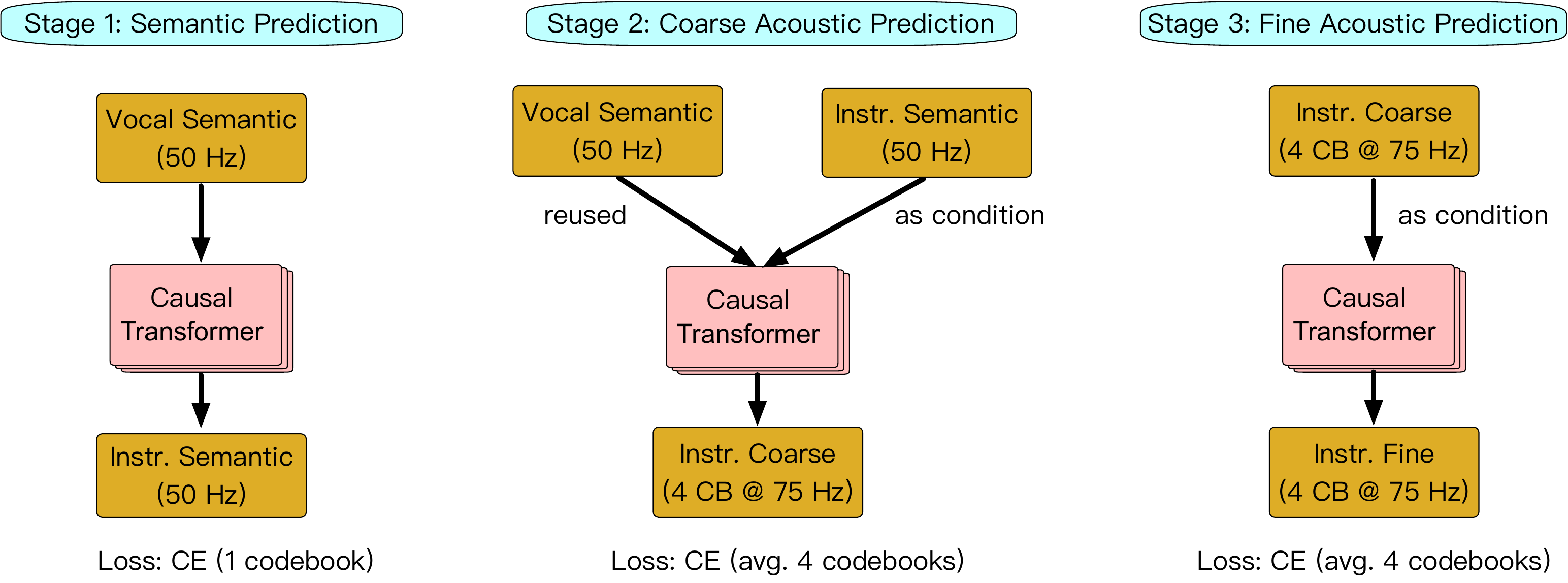}
\caption{Training procedure for the three-stage hierarchical autoregressive model. Each stage is trained independently with teacher forcing and CFG dropout ($p_{\text{drop}}\!=\!0.1$). Conditioning and target tokens are embedded with separate token and segment embeddings, concatenated, and fed through a causal Transformer. Per-quantizer linear heads produce output logits.}
\label{fig:training}

\end{figure*}

\section{Related Work}
\label{sec:related}

\textbf{Audio-domain accompaniment generation.}
SingSong~\cite{donahue2023singsong} is the most closely related work, adapting AudioLM~\cite{borsos2022audiolm} for conditional audio-to-audio generation. It uses source separation to create training pairs, encodes vocals with w2v-BERT~\cite{chung2021w2vbert}, and models instrumentals via SoundStream~\cite{zeghidour2021soundstream} tokens using a T5 encoder-decoder. Key to its generalization is adding noise to vocals and using only semantic codes (S-SA featurization). Our work builds on these insights but replaces the encoder-decoder with a decoder-only AR approach and upgrades the codec pipeline.

\textbf{Neural audio generation.}
AudioLM~\cite{borsos2022audiolm} introduced hierarchical generation of semantic and acoustic codes for unconditional audio synthesis. MusicLM~\cite{agostinelli2023musiclm} extended this to text-conditioned music generation. VALL-E~\cite{wang2023neural} applied similar hierarchical AR modeling to speech synthesis. Our three-stage decomposition follows this paradigm but targets the cross-modal vocal-to-instrumental task.

\textbf{Neural audio codecs.}
SoundStream~\cite{zeghidour2021soundstream} and EnCodec~\cite{defossez2022encodec} use residual vector quantization (RVQ) to compress audio into discrete tokens. EnCodec offers higher quality at comparable bitrates and supports flexible bandwidth configurations (1.5--24\,kbps). We adopt EnCodec at 6.0\,kbps (8 codebooks, 75\,Hz) for instrumental tokenization.

\section{The Proposed Methodology}
The HAFM system consists of three main components: data preprocessing and tokenization, three-stage hierarchical autoregressive generation, and decoding with audio reconstruction. As illustrated in Fig.~\ref{fig:preprocess}, the system first encodes the vocals and instrumentals separately into token sequences at different rates, then progressively generates instrumental tokens through a three-stage cascaded Transformer, and finally decodes them into audio waveforms. Each component is described in detail below.

\subsection{Data Preprocessing and Tokenization}
\label{section:proposed_method} 
The data preprocessing pipeline is illustrated in Figure 1. First, the MDXNet~\cite{kim2021kuielab} source separation model is used to extract aligned vocal-instrumental pairs from music mixtures. To enhance the model's robustness to noise, Gaussian noise ( = 0.01) is applied to the vocals as data augmentation. For vocal tokenization, the HuBERT-Base model is used to extract semantic features, which are then quantized into discrete tokens via k-means clustering (k = 500) at a sampling rate of 50 Hz. For instrumental tokenization, EnCodec (24 kHz, 8 codebooks) is used to encode the instrumentals into acoustic tokens at a sampling rate of 75 Hz. The 8 codebooks are divided into two groups—coarse-grained (the first 4, codebooks 0–3) and fine-grained (the last 4, codebooks 4–7)—for use in hierarchical generation.

\subsection{Three-Stage Hierarchical Architecture}
Inspired by AudioLM and MusicLM, HAFM is optimized for the accompaniment generation task and adopts a three-stage hierarchical generation architecture that progressively refines audio quality. The three stages are as follows:

(1) Stage 1: Semantic Modeling. This stage targets the HuBERT semantic token sequence and learns the high-level semantic structure of music. The model takes the HuBERT tokens of the vocals as input and outputs the HuBERT tokens of the accompaniment. This stage captures macroscopic musical features such as melody, harmony, and rhythm.

(2) Stage 2: Coarse-Grained Acoustic Modeling. This stage targets the first 4 quantization layers of EnCodec (Q0–Q3), adding acoustic detail on top of the semantic tokens. The model takes as input all EnCodec tokens of the vocals along with the accompaniment HuBERT tokens generated in Stage 1, and outputs the coarse-grained EnCodec tokens of the accompaniment. This stage begins to introduce acoustic features such as timbre and loudness.

(3) Stage 3: Fine-Grained Acoustic Modeling. This stage targets the last 4 quantization layers of EnCodec (Q4–Q7), further refining audio quality. The model takes as input all EnCodec tokens of the vocals, the HuBERT tokens generated in Stage 1, and the coarse-grained EnCodec tokens generated in Stage 2, and outputs the fine-grained EnCodec tokens of the accompaniment. This stage adds high-frequency detail and subtle acoustic textures, ultimately producing high-fidelity audio.

\subsection{Classifier-Free Guidance}

To strengthen the model's conditional dependence on the input vocals, HAFM employs Classifier-Free Guidance (CFG). CFG was originally developed for diffusion models; in this work, it is adapted for autoregressive language models.

During training, the conditional input (i.e., vocal tokens) is randomly dropped with a probability of $10\%$, allowing the model to learn both unconditional and conditional generation modes. During inference, the output logits are adjusted according to the following formula:
\begin{equation}
    \text{logits} = \text{logits}_u + \lambda \times (\text{logits}_c - \text{logits}_u) 
\end{equation}

where $\text{logits}_c$ denotes the output under conditional generation, $\text{logits}_u$ denotes the output under unconditional generation, and $\lambda$ is the guidance strength hyperparameter. A larger value of $\lambda$ increases the correlation between the generated output and the input vocals, but may reduce diversity. In our experiments, $\lambda$ is set to 3.0.

\subsection{Transformer Architecture Design}

HAFM adopts a modern Transformer architecture that incorporates several recent improvements to enhance training stability and generation quality. The model is configured with 12 layers, a hidden dimension of 512, 8 attention heads, and a feed-forward network dimension of 2048. The main architectural features are as follows:

(1) QK Normalization: The Query and Key vectors in the attention mechanism are normalized to prevent gradient vanishing caused by excessively large dot-product values, thereby improving training stability.

(2) GEGLU Activation Function: A variant of the Gated Linear Unit (GLU) is used in the feed-forward network, offering greater expressive capacity compared to the conventional ReLU activation function.

(3) RMSNorm: Root Mean Square Layer Normalization is used in place of conventional LayerNorm, achieving comparable performance with higher computational efficiency.

(4) T5 Relative Position Bias: The relative position encoding scheme from the T5 model is adopted, enabling the model to better capture relative positional relationships within sequences. This is particularly important for music, which exhibits strong temporal structure.

\begin{figure}
  \centering
  \includegraphics[width=8cm]{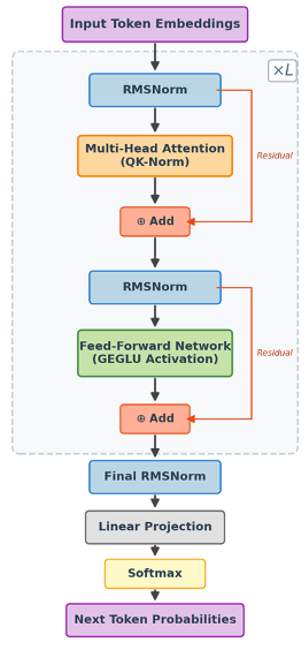}
  \caption{Transformer Module Architecture.}
  \label{fig:03}
\end{figure}

Fig.~\ref{fig:03} illustrates the detailed structure of a single Transformer layer. Each layer contains a multi-head self-attention module and a feed-forward network module, both employing residual connections and RMSNorm. The attention module uses a causal mask to ensure the autoregressive generation property. These modern architectural improvements enable HAFM to train efficiently and generate high-quality music accompaniment.

\subsection{Model Training}
This section describes in detail the training process of the three-stage HAFM model, including dataset preparation, training configuration, and evaluation metrics.

The model is trained on the FMA (Free Music Archive)~\cite{defferrard2016fma}, which contains 106,574 complete musical works spanning a wide variety of musical styles. The preprocessing pipeline applies MDXNet for source separation, splitting each song into vocal and instrumental tracks, which are then segmented into 10-second clips. After quality filtering, a final set of 44,547 training samples and 5,497 validation samples is obtained. For each sample, HuBERT is used to extract 50 Hz vocal semantic tokens (vocal\_semantic), and EnCodec is used to extract 75 Hz instrumental semantic tokens (inst\_semantic) and acoustic tokens (inst\_coarse for the first 4 layers and inst\_fine for the last 4 layers).

Three-Stage Training Configuration: The model adopts a cascaded training strategy, training the semantic, coarse acoustic, and fine acoustic stages sequentially.

Semantic Stage: The model contains 38.6M parameters. The input consists of vocal semantic tokens (50 Hz) and the output consists of instrumental semantic tokens (50 Hz). Training uses a batch size of 36 and runs for approximately 36K steps on 2 A40 GPUs using DDP. The best cross-entropy loss on the validation set is 2.4441, with a token-level accuracy of $46\%$. This stage learns the high-level semantic mapping from vocals to instrumentals.

Coarse Stage: The model contains 42.5M parameters. The input consists of vocal semantic tokens and instrumental semantic tokens, and the output consists of coarse-grained instrumental acoustic tokens (the first 4 codebooks, 75 Hz). Training uses a batch size of 12 and runs for approximately 354K steps. The best validation loss is 3.9029, with an accuracy of approximately $20\%$. This stage converts semantic information into coarse-grained acoustic representations, establishing fundamental acoustic structures such as pitch and rhythm.

Fine Stage: The model contains 44.1M parameters. The input consists of coarse instrumental acoustic tokens (first 4 layers) and the output consists of fine-grained instrumental acoustic tokens (last 4 layers, 75 Hz). Training uses a batch size of 36 and runs for 400K steps. The best validation loss is 4.7696, with an accuracy of approximately $10\%$. This stage is responsible for recovering high-frequency acoustic details such as timbre and reverberation.

All stages use the AdamW optimizer~\cite{kingma2014adam} (learning rate 1e-4, weight decay 0.01) with FP16 mixed-precision training to improve efficiency. Classifier-Free Guidance (CFG) is applied throughout training, randomly dropping conditional inputs with a probability of $10\%$; during inference, a guidance strength of 3.0 is used to enhance conditional control. 

Evaluation Metrics: Two metrics are used to monitor model performance during training: (1) **Cross-entropy loss:** measures the discrepancy between the predicted token distribution and the ground-truth distribution; lower values indicate better model fit. (2) **Token-level accuracy:** computes the proportion of predicted tokens that exactly match the ground-truth tokens, i.e., accuracy = (number of correctly predicted tokens) / (total number of tokens). Since later stages operate over larger codebook vocabularies with richer acoustic detail, accuracy decreases progressively across stages (Semantic: $46\%$, Coarse: $20\%$, Fine: $10\%$), which is an expected and normal phenomenon.

\section{Experiments}
\subsection{Comparative Experiments}
\subsubsection{Dataset and Evaluation Tasks}

This paper constructs an accompaniment generation evaluation set based on the MUSDB18~\cite{rafii2017musdb18} test set. Specifically, 33 audio clips are selected from the test set, each with a duration of 3 seconds, and a pure vocal reference file prompt\_vocal.wav is retained for each clip. Around the same vocal segment, four categories of candidate system outputs are constructed: the generated model output (model), the ground-truth accompaniment (gt), and a random baseline (random). Here, model refers to the accompaniment generated by the proposed method, gt refers to the corresponding ground-truth accompaniment from the dataset, and random refers to a randomly selected accompaniment clip.

To more rigorously compare the degree of vocal matching across different systems, this paper adopts a pairwise comparison evaluation protocol. For each clip, all pairwise combinations are formed from the 3 systems, yielding $\binom{3}{2} = 3$ comparison pairs per clip: model vs. gt, model vs. random, and gt vs. random. Consequently, each model version corresponds to $33 \times 3 = 99$ pairwise comparison combinations.

\subsubsection{Listening Test Design}

The listening test adopts an A/B preference test format. Each trial consists of three audio clips: first, prompt\_vocal.wav is played, which contains only the pure vocal; this is followed by A.wav and B.wav, both of which are vocal-plus-accompaniment mixtures derived from two different systems under comparison. After listening to all three audio clips, evaluators are asked to judge: given the provided vocal, which accompaniment—A or B—better matches the vocal.

The evaluation criteria cover three main dimensions: (1) **Rhythmic matching**, i.e., whether the beats and groove of the accompaniment align naturally with the vocal; (2) **Harmonic matching**, i.e., whether the harmonic progressions of the accompaniment are consonant with the melody; and (3) **Overall musical coherence**, i.e., whether the accompaniment sounds as though it was specifically tailored for the given vocal segment. The subjective evaluation does not primarily judge audio quality in isolation, but rather focuses on the degree of match between the accompaniment and the vocal. To avoid prior bias, all pairwise trials are conducted under a blind evaluation setting, in which the true system identities corresponding to A and B are randomly shuffled; the mapping is recorded and the true labels are only restored during the statistical analysis stage.

\subsubsection{Automatic Evaluation Design}
This paper further introduces the audio multimodal large language model Qwen2-Audio-7B-Instruct~\cite{chu2024qwen2} as an automatic evaluator to score the same set of pairwise trials. In the specific implementation, for each trial, vocal.wav, A.wav, and B.wav are sequentially fed into Qwen2-Audio, and a unified prompt instructs the model to determine which of A or B is superior across three dimensions: rhythmic alignment, harmonic compatibility, and overall musical coherence. The model output comprises three components: the selection result, a confidence score, and a brief reasoning explanation.

On the engineering side, two A40 GPUs are used in parallel to execute the automatic evaluation, with the 99 trials evenly distributed across the two GPUs for separate processing; results are merged and metrics are computed after completion. The evaluation script primarily outputs three categories of results: (1) the overall number of wins and win rate for each system; (2) head-to-head win rates for each pairwise comparison; and (3) confidence-weighted win rates, along with fine-grained analysis aggregated at the clip level.

\subsubsection{Comparative Experiment Results}

\begin{table}[h]
\centering
\caption{Overall Win Rate}
\label{tab:overall_win_rate}
\begin{tabular}{lccc}
\toprule
\textbf{System} & \textbf{Wins} & \textbf{Total Matches} & \textbf{Win Rate} \\
\midrule
model  & 36 & 99 & 36.4\% \\
gt     & 34 & 99 & 34.3\% \\
random & 29 & 99 & 29.3\% \\
\bottomrule
\end{tabular}
\end{table}

The Qwen2-Audio automatic evaluation results show that under the BASE model configuration, the overall win rates of the three systems are as follows: model $36.4\%$, gt $34.3\%$, and random $29.3\%$. In terms of overall ranking, the proposed generative model (model) ranks first, outperforming both the ground-truth baseline and the random baseline, indicating that the model is already capable of generating accompaniments with a meaningful degree of vocal matching. Examining the direct pairwise comparisons of the generative model against other systems, in the BASE version, the win rate of model vs. gt is $51.5\%$ (17/33), and model vs. random is $57.6\%$ (19/33).

\begin{table}[h]
\centering
\caption{Head-to-Head Comparison Results}
\label{tab:head_to_head}
\begin{tabular}{lcc}
\toprule
\textbf{Comparison} & \textbf{Wins/Total} & \textbf{Win Rate} \\
\midrule
Model vs. gt     & 17/33 & 51.5\% \\
Model vs. random & 19/33 & 57.6\% \\
Gt vs. random    & 18/33 & 54.5\% \\
\bottomrule
\end{tabular}
\end{table}

These results demonstrate that the proposed generative model is already able to achieve a win rate slightly above $50\%$ in direct comparisons with ground-truth accompaniments, meaning that its generated outputs are not only usable on certain samples, but can approach or even surpass the ground truth in terms of vocal matching. On the other hand, the generative model maintains a clear advantage over the random baseline, indicating that the model has genuinely learned effective vocal-accompaniment correspondences rather than merely outputting a statistically averaged pattern.

\begin{table}[h]
\centering
\caption{Confidence-Weighted Results}
\label{tab:confidence_weighted}
\begin{tabular}{lcc}
\toprule
\textbf{System} & \textbf{Weighted Score/Total Confidence} & \textbf{Weighted Win Rate} \\
\midrule
model  & 150/434 & 34.56\% \\
gt     & 148/434 & 34.10\% \\
random & 136/434 & 31.34\% \\
\bottomrule
\end{tabular}
\end{table}

To examine the stability of the automatic evaluation results, this paper further computes confidence-weighted win rates based on the model's confidence scores. The results show that in the BASE version, the confidence-weighted win rates of model, gt, and random are $34.56\%$, $34.10\%$, and $31.34\%$, respectively. It can be observed that the system ranking after weighted aggregation is largely consistent with the unweighted results, indicating that Qwen2-Audio's judgments exhibit good internal consistency and are not driven by a small number of low-confidence samples.

\subsubsection{Results Analysis}
First, the generative model demonstrates a consistent advantage over the random baseline and achieves a win rate of $51.5\%$ in head-to-head comparisons against the ground-truth accompaniment. This indicates that the model does not merely generate signals that sound like accompaniment, but has to some extent learned to organize appropriate rhythm and harmony around the vocal melody. From the perspective of the accompaniment generation task, this is a meaningful result, as it suggests that the model's optimization objective is genuinely aligned with the core problem of vocal-accompaniment matching. Second, the overall win rate distribution across the three systems is relatively close, falling roughly within the range of $29.3\%$ to $36.4\%$, indicating that the task itself is inherently challenging and that no system holds an overwhelmingly dominant advantage over the others. From another perspective, this also implies that the proposed generative model has already approached the level of ground-truth accompaniment and is highly competitive.

Taking the subjective A/B preference evaluation framework and the Qwen2-Audio automatic evaluation results together, it can be seen that the accompaniment generated by the proposed method is clearly superior to the random baseline in terms of rhythmic matching with the vocals, harmonic compatibility, and overall musical coherence, and achieves matching quality that equals or even surpasses the ground-truth accompaniment on a subset of samples. The proposed generative model has already demonstrated strong vocal-conditioned modeling capability.

\subsection{Ablation Study}
To validate the effectiveness of the HAFM three-stage architecture, this section designs an ablation study comparing the impact of different stage combinations on generation quality. The experiment uses FAD (Fr{\'e}chet Audio Distance)~\cite{kilgour1812frechet} as the objective evaluation metric, evaluated on 33 samples from the MUSDB18 test set.

Experimental Setup: Two sets of comparative experiments are designed.

(1) Baseline (Semantic + Coarse): Only the first two stages are used, generating coarse-grained audio with 4 codebooks. This configuration skips the Fine stage and directly decodes the first 4 acoustic code layers output by the Coarse stage into waveforms via EnCodec.

\begin{table}[h]
\centering
\caption{Ablation Study Results}
\label{tab:ablation}
\begin{tabular}{lccc}
\toprule
\textbf{Model Configuration} & \textbf{Stages} & \textbf{Codebooks} & \textbf{FAD↓} \\
\midrule
Baseline        & Semantic + Coarse            & 4 & 2.10 \\
Full (proposed) & Semantic + Coarse + Fine     & 8 & 1.71 \\
\midrule
Improvement     & —                            & — & 0.39 (18.6\%) \\
\bottomrule
\end{tabular}
\end{table}

(2) Full (Semantic + Coarse + Fine): The complete three-stage pipeline is used, generating high-quality audio with 8 codebooks. This configuration sequentially executes semantic generation, coarse acoustic generation, and fine acoustic generation, ultimately outputting the complete 8-layer acoustic codes.

Both experimental groups use the same test set, the same generation parameters (temperature = 0.9, $top\_k$ = 250, $cfg\_scale$ = 3.0), and the same evaluation pipeline to ensure a fair comparison. Table 4 presents the quantitative results of the ablation study. The results show that adding the Fine stage reduces FAD from 2.10 to 1.71, a relative improvement of $18.6\%$. This significant improvement demonstrates the importance of the Fine stage in recovering high-frequency acoustic details. Specifically:

(1) Although the Baseline model is capable of generating basic pitch and rhythmic structures, the use of only 4 codebooks results in thin timbre and a lack of detail, leading to a relatively large distributional gap between the generated and real audio.

(2) The Full model supplements the last 4 codebooks through the Fine stage, significantly enhancing timbral richness, reverberation, and high-frequency detail, bringing the generated audio closer to the acoustic characteristics of real accompaniment.

(3) The $18.6\%$ improvement in FAD indicates that the hierarchical modeling strategy (coarse-to-fine) holds a clear advantage over single-stage generation, validating the rationale behind the proposed three-stage architectural design.

In terms of subjective listening impression, the accompaniment generated by the Baseline sounds relatively dry and lacks the texture of real instruments, whereas the accompaniment generated by the Full model is rich in timbre and detail, with a higher degree of integration with the vocals. This is consistent with the objective evaluation results from the FAD metric.

\subsection{Model Convergence Validation}
o further validate the effectiveness and optimization stability of the HAFM three-stage training strategy, Figures 4 through 6 present the training and validation convergence curves for the semantic stage, the coarse-grained acoustic stage, and the fine-grained acoustic stage, respectively. In each figure, the left axis represents the loss value, the right axis represents accuracy, solid lines indicate training set metrics, and token-marked curves indicate validation set metrics.
\begin{figure}
  \centering
  \includegraphics[width=9.5cm]{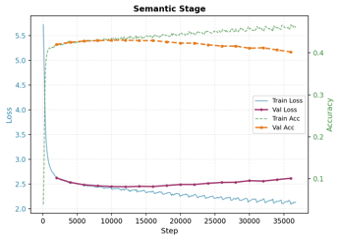}
  \caption{Semantic-Stage Convergence Curve.}
  \label{fig:04}
\end{figure}

As shown in Fig.~\ref{fig:04}, the training loss and validation loss of the semantic stage decline consistently throughout training, while accuracy rises steadily, indicating that the model is able to quickly learn the melodic, rhythmic, and harmonic structures corresponding to the input vocals. The training and validation curves follow broadly consistent trends without notable divergence, demonstrating good generalization capability at this stage.
\begin{figure}
  \centering
  \includegraphics[width=9.5cm]{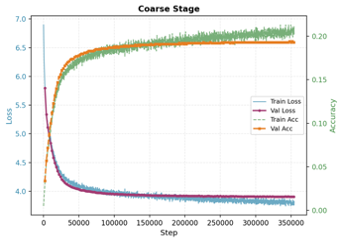}
  \caption{Coarse-Grained Acoustic-Stage Convergence Curve.}
  \label{fig:05}
\end{figure}

Fig.~\ref{fig:05} shows that the coarse-grained acoustic stage maintains stable convergence over a longer training process. As the number of training steps increases, the loss continues to decrease and the validation accuracy improves in tandem, indicating that under the conditional constraints of the semantic tokens, the model is able to continuously learn the primary timbre, rhythmic energy distribution, and time-frequency structure of the instrumental accompaniment.
\begin{figure}
  \centering
  \includegraphics[width=9.5cm]{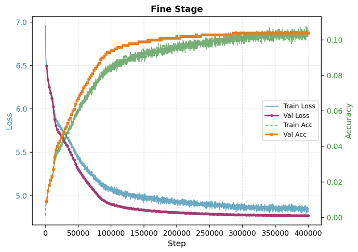}
  \caption{Fine-Grained Acoustic-Stage Convergence Curve.}
  \label{fig:06}
\end{figure}

As shown in Fig.~\ref{fig:06}, the fine-grained acoustic stage converges at a relatively slower pace, though the overall optimization trend remains stable. Given the presence of a small number of anomalous fluctuations in the later stages of training, the figure shows only the convergence results from the stable training interval. After filtering out the anomalous points, both the training loss and validation loss maintain a downward trend with accuracy improving in parallel, indicating that the model is able to further supplement high-frequency details and textural information without disrupting the overall structure established by the preceding two stages.

In summary, all three stages of the model exhibit good optimization stability and consistent training-validation trends, validating from the perspective of the training process that the proposed hierarchical autoregressive modeling framework possesses strong trainability and convergence. This result also provides direct support for the performance improvements observed in subsequent objective metrics and subjective listening evaluations.

\subsection{Discussion}
The success of HAFM validates the effectiveness of hierarchical autoregressive modeling for the music accompaniment generation task. By decomposing the generation process into three stages—semantic, coarse acoustic, and fine acoustic—the model is able to capture musical structure at different levels of abstraction, ensuring both semantic coherence and high-fidelity acoustic detail. This design philosophy also offers valuable insights for other audio generation tasks.

The dual-rate encoding scheme represents another key innovation. HuBERT's semantic tokens (50 Hz) capture the high-level structure of music, while EnCodec's acoustic tokens (75 Hz) encode fine-grained audio details. The combination of the two enables the model to simultaneously model both semantic and acoustic dimensions, avoiding the limitations of a single encoding scheme. This multimodal encoding approach is worth further exploration in future audio generation research.

The application of classifier-free guidance within the autoregressive model has also yielded notable results. By randomly dropping conditional inputs during training, the model learns to distinguish between conditional and unconditional generation; during inference, the conditional dependence is enhanced by interpolating between the two. This technique is simple yet effective, requiring no additional classifier training, and is particularly well-suited to tasks that demand strong conditional control, such as music accompaniment generation.

The current model also has certain limitations. First, the three-stage training requires considerable training time and computational resources. Future work could explore end-to-end joint training schemes to improve training efficiency. Second, the model is currently trained only on the FMA Large dataset, and its generalization ability warrants further validation. Training on larger and more diverse music datasets may further improve performance. Finally, while the accompaniment generated by the model is of relatively high quality, there remains room for improvement on certain complex musical styles.

\section{Conclusion and Future Work}
This paper presents HAFM — a hierarchical autoregressive foundation model designed for the music accompaniment generation task. Targeting the shortcomings of existing methods in three areas — high-level semantic modeling, acoustic detail reconstruction, and conditional control — HAFM introduces systematic improvements across three dimensions: representation learning, generative architecture, and training strategy.

At the representation level, the model adopts a dual-rate tokenization scheme that combines 50 Hz HuBERT semantic tokens with 75 Hz EnCodec acoustic tokens, achieving decoupled modeling of semantic and acoustic information and laying the foundation for subsequent hierarchical generation. At the architectural level, the three-stage cascaded generation framework decomposes accompaniment generation into three subtasks — semantic generation, coarse-grained acoustic generation, and fine-grained acoustic generation — enabling the model to progressively refine musical structure and acoustic detail at different levels of abstraction. At the training strategy level, the classifier-free guidance mechanism effectively strengthens the model's conditional dependence on the input vocals; combined with modern Transformer components including QK normalization, GEGLU activations, RMSNorm, and T5 relative position bias, training stability and generation quality are further enhanced.

Experiments on the MUSDB18 dataset demonstrate that the complete three-stage HAFM model achieves an FAD score of 1.71, representing an $18.6\%$ improvement over the two-stage baseline. In subjective listening evaluations, the generated accompaniment achieves a win rate of $51.5\%$ in direct comparisons against ground-truth accompaniment, demonstrating strong competitiveness in rhythmic matching, harmonic coherence, and overall musical fit. Ablation studies validate the effectiveness of each core component from both objective metric and convergence curve perspectives, confirming the feasibility and superiority of the hierarchical autoregressive modeling framework for this task.

Looking ahead, this paper identifies the following directions for future research: first, exploring end-to-end joint training schemes for the three-stage model to reduce cascaded error accumulation and improve overall training efficiency; second, expanding the scale and stylistic diversity of training data to validate and enhance the model's generalization ability across different musical genres such as pop, classical, and jazz; third, introducing finer-grained generation control mechanisms to support user-specified high-level attributes such as instrumentation, musical style, or emotional characteristics, thereby improving system interactivity; fourth, investigating lightweight inference solutions for real-time scenarios and exploring the application of HAFM to latency-sensitive practical settings such as live music performance accompaniment assistance.

HAFM provides a systematic solution for vocal-conditioned music accompaniment generation, and its hierarchical modeling approach and dual-rate tokenization strategy offer meaningful reference value for the broader field of audio generation research.

\section*{Acknowledgment}
This work is supported by the National Key Research and Development Program of China (Grant No. 2021ZD0201501), the National Natural Science Foundation of China (No. 22574146, No. 32200860, and No.62306289), by the Regional Innovation and Development Joint Fund of National Natural Science Foundation of China (U22A6001).

\bibliography{main}

\end{document}